# On Concurrent and Resettable Zero-Knowledge Proofs for NP[*]

Joe Kilian [†]    Erez Petrank [‡]    Ransom Richardson [§]

## Abstract

A proof is concurrent zero-knowledge if it remains zero-knowledge when many copies of the proof are run in an asynchronous environment, such as the Internet. It is known that zero-knowledge is not necessarily preserved in such an environment [24, 29, 4]. Designing concurrent zero-knowledge proofs is a fundamental issue in the study of zero-knowledge since known zero-knowledge protocols cannot be run in a realistic modern computing environment. In this paper we present a concurrent zero-knowledge proof systems for all languages in NP. Currently, the proof system we present is the only known proof system that retains the zero-knowledge property when copies of the proof are allowed to run in an asynchronous environment. Our proof system has $\tilde{O}(\log^2 k)$ rounds (for a security parameter $k$), which is almost optimal, as it is shown in [4] that black-box concurrent zero-knowledge requires $\tilde{\Omega}(\log k)$ rounds.

Canetti, Goldreich, Goldwasser and Micali introduced the notion of *resettable* zero-knowledge, and modified an earlier version of our proof system to obtain the first resettable zero-knowledge proof system. This protocol requires $k^{\theta(1)}$ rounds. We note that their technique also applies to our current proof system, yielding a resettable zero-knowledge proof for NP with $\tilde{O}(\log^2 k)$ rounds.

## 1 Introduction

Zero-knowledge proof systems, introduced by Goldwasser Micali and Rackoff [21], are efficient interactive proofs that yield no knowledge but the validity of the proven assertion. These proofs have proven important tools for a variety of cryptographic applications. However, the original definition of zero-knowledge considers security only in a restricted scenario in which the prover and the verifier execute one instance of the proof disconnected from the rest of the computing environment.

In recent years, several papers have studied the affect of a modern computing environment on the security of zero-knowledge. In particular, many computers today are connected through networks in which connections are maintained in parallel asynchronous sessions. It would be common to find several connections (such as FTP, Telnet, An internet browser, etc.) running together on a single workstation. Can zero-knowledge protocols be trusted in such an environment?

---

[*]This paper is a join of two works. The preliminary versions of these works appeared in the Proceeedings of *Advances in Cryptology - EUROCRYPT '99*, May 1999, Lecture Notes in Computer Science Vol. 1592 Springer 1999, pp. 415-431, and in the *Proceedings of the thirty third annual ACM Symposium on Theory of Computing*. ACM Press, 2001.

[†]Yianilos Labs, joe@pnylab.com. Work done in part while at the NEC Research Institute.

[‡]Dept. of Computer Science, Technion - Israel Institute of Technology, Haifa 32000, Israel. erez@cs.technion.ac.il. This research was supported by Technion V.P.R. Fund - N. Haar and R. Zinn Research Fund.

[§]Groove Networks, 800 Cummings Center, Beverly, MA 01915. rrichard@groove.net.

## 1.1 Previous work

Composing zero-knowledge proofs in asynchronous environment was first mentioned by Feige [11], and first explored in a rigorous setting by Dwork, Naor, and Sahai [9]. Dwork, Naor and Sahai denoted zero-knowledge protocols that are robust to asynchronous composition *concurrent zero-knowledge* protocols. They observed that several known zero-knowledge proofs, with a straight-forward adaptation of their original simulation to the asynchronous environment, may cause the simulator to work exponential time. Thus, it seems that the zero-knowledge property does not necessarily carry over to the asynchronous setting.

Kilian, Petrank, and Rackoff [24] gave the first lower bound for concurrent zero-knowledge, showing that any language that has a 4-rounds concurrent (black-box) zero-knowledge interactive proof or argument is in BPP. Thus, a large class of known zero-knowledge interactive proofs and arguments for non-trivial languages do not remain zero-knowledge in an asynchronous environment. Rosen [29] has improved this lower bound from from 4 rounds to 7, and Canetti, Kilian, Petrank and Rosen have recently improved the lower bound substantially showing that concurrent black-box zero-knowledge proof systems for non-trivial languages require $\tilde{\Omega}(\log k)$ rounds. A natural question is whether there exists a fully asynchronous (concurrent) zero-knowledge proof for NP.

## 1.2 This work

In this paper, we exhibit the first (and currently the only known) concurrent (black-box) zero-knowledge interactive proof for any language in NP. Our proof system has $\omega(\log^2 k)$ rounds, where $k$ is the security parameter. Namely, the running time of all efficient parties is bounded by a polynomial in $k$, and so is the number of copies of the proof that may be run concurrently. Saying that the number of rounds is $\omega(\log^2 k)$ rounds we mean that the number of rounds may be set to $h(k) \cdot (\log^2 k)$, for any function $h : \mathcal{N} \to \mathcal{N}$ that cannot be bounded by any constant.

The concurrent zero-knowledge interactive proof for NP that we present relies on the existence of secure bit commitment schemes with statistical binding and bit commitment schemes with statistical secrecy (see Section 2.2 below). Using [25] the first can be efficiently based on any pseudo-random generator (and thus on any one-way function), and using [1, 26, 7], the latter can be efficiently based on the existence of collision-intractable hash functions. Our proof system may be run by an efficient prover that is given the witness for the (NP) assertion in the input.

Finally, the proof system we present is a family of proofs systems. It is parameterized by the number of rounds. The proof as presented may be used with any number of rounds (above some required constant). The analysis we provide shows that it is concurrent zero-knowledge when the number of rounds is set to $\omega(\log^2 k)$. By the lower bound, we know that if we set the number of rounds too low, then the proof system does not remain concurrent zero-knowledge. Possibly, the same proof system remains zero-knowledge even if the number of rounds is set to $O(\log n)$, but we do not know how to show it. This is an interesting open question.

### 1.2.1 Resettable zero-knowledge

Finally, we note that the techniques in [3] apply to our new protocol. Thus, our concurrent zero-knowledge interactive proof can be modified to be made *resettable* zero-knowledge. Resettable zero-knowledge proofs were presented by Canetti, Goldreich, Goldwasser and Micali [3]. Such proofs are zero-knowledge proofs that on top of being concurrent, maintain zero-knowledge properties when the verifier is allowed to run the prover repeatedly on a fixed (yet, randomly chosen) random tape.

The practical motivation behind such robustness is a use of zero-knowledge in smartcards, where the prover (the card) can be reset by the verifier to run repeatedly without access to additional random coin tosses. It is assumed that the actual random tape of the card is hidden from the verifier, and it is shown in [3] how this hidden randomness can be used to allow such robustness of a zero-knowledge proof.

Canetti, Goldreich, Goldwasser and Micali modified an earlier version of our proof system to obtain the first resettable zero-knowledge proof system. This protocol requires $k^{\theta(1)}$ rounds. We note that their technique also applies to our current proof system, yielding a resettable zero-knowledge proof for NP with $\tilde{O}(\log^2 k)$ rounds. Previously, there was no sub-polynomial resettable zero-knowledge proof in a general asynchronous environment. We remark that as with concurrent zero-knowledge, if one makes set-up assumptions, then one may get more efficient proofs. For example, in the *public key model* there exist more efficient resettable zero-knowledge proofs, see [3].

The definition of resettable zero-knowledge and related issues appear in Section 8.1 below. For a more detailed discussion, motivation, and definitions the reader is referred to [3].

### 1.3 In light of the lower bound

Several works have overcome the difficulty of the asynchronous setting by using some compromises. For example, compromising the strength of zero-knowledge security, introducing *witness indistinguishability* [14], or putting limits on the asynchronisity of the system (a.k.a timing assumptions) [9, 10, 5], or by making some set-up assumptions on the environment (such as a public key infrastructure) [6, 3].

### 1.4 Terminology

Some words on the terminology we are using. By zero-knowledge we mean *computational* zero-knowledge, i.e., the distribution output by the simulation is polynomial-time indistinguishable from the distribution of the views of the verifier in the original interaction. (See definitions in Section 2.1 below.) Our proof is black-box zero-knowledge (see Section 2.4 below). The proof will be perfectly sound, i.e., we will construct an interactive proof, yet it will be possible to run the prover in polynomial time given a witness to the NP assertion that the prover is making.

### 1.5 Guide to the paper

In Section 2 we present some definitions and the tools we are using. In Section 3 we state our results. In Section 4 we present the concurrent zero-knowledge interactive proof for NP. In Section 5 we provide a simulator for the interaction between the prover and any (adversarial) verifier. In Section 6 we analyze the simulator with respect to a static schedule. Namely, the schedule may be the worst possible, but it is not modified during the rewinds of the simulator. In Section 7 we show that the simulator works as well also with respect to schedules that change dynamically during the simulation. Thus, our proof system is concurrent zero-knowledge. Finally, in Section 8, we discuss our protocol in the resettable zero-knowledge model.

## 2 Preliminaries

In this section we go over the definitions and the tools we are using. We postpone the definitions and discussion of resettable zero-knowledge to Section 8.

### 2.1 Zero-knowledge proofs

Let us recall the concept of interactive proofs, as presented by [21]. For formal definitions and motivating discussions the reader is referred to [21].

**Definition 2.1** *A protocol between a (computationally unbounded)* prover *$P$ and a (probabilistic polynomial-time)* verifier *$V$ constitutes an* interactive proof *for a language $L$ if there exists a negligible fraction $\varepsilon$ such that*

- **Completeness:** *If $x \in L$ then*
$$\Pr\left[(P,V)(x) \text{ accepts}\right] \geq 1 - \varepsilon(|x|)$$

- **Soundness:** *If $x \notin L$ then for any prover $P^*$*
$$\Pr\left[(P^*,V)(x) \text{ accepts}\right] \leq \varepsilon(|x|)$$

Brassard, Chaum, and Crépeau [1] suggested a modification of interactive proofs called *arguments* in which the prover is also polynomial time bounded. Thus, the soundness property is modified to be guaranteed only for probabilistic polynomial time provers $P^*$.

Let $(P,V)(x)$ denote the random variable that represents $V$'s view of the interaction with $P$ on common input $x$. The view contains the verifier's random tape as well as the sequence of messages exchanged between the parties.

We briefly recall the definition of black-box zero-knowledge [21, 27, 17, 20]. The reader is referred to [20] for more details and motivation.

**Definition 2.2** *A protocol $(P,V)$ is* computational zero-knowledge *(resp.,* statistical zero-knowledge*) over a language $L$, if there exists an oracle probabilistic polynomial time machine $S$ (simulator) such that for any polynomial time verifier $V^*$ and for every $x \in L$, the distribution of the random variable $S^{V^*}(x)$ is polynomially indistinguishable from the distribution of the random variable $(P,V^*)(x)$ (resp., the statistical difference between $M(x)$ and $(P,V)(x)$ is a negligible function in $|x|$).*

In this paper, we concentrate on computational zero-knowledge. In the sequel we will say *zero-knowledge* meaning *computational zero-knowledge*.

## 2.2 Bit commitments

We include a short and informal presentation of commitment schemes. For more details and motivation, see [15]. A commitment scheme involves two parties: The *sender* and the *receiver*. These two parties are involved in a protocol which contains two phases. In the first phase the sender commits to a bit, and in the second phase it reveals it. A useful intuition to keep in mind is the "envelope implementation" of bit commitment. In this implementation, the sender writes a bit on a piece of paper, puts it in an envelope and gives the envelope to the receiver. In a second (later) phase, the *reveal* phase, the receiver opens the envelope to discover the bit that was committed on. In the actual digital protocol, we cannot use envelopes, but the goal of the cryptographic machinery used, is to simulate this process.

More formally, a commitment scheme consists of two phases. First comes the *commit* phase and then we have the *reveal* phase. We make two security requirements which (loosely speaking) are:

**Secrecy:** At the end of the *commit phase*, the receiver has no knowledge about the value committed upon.

**Binding property:** It is infeasible for the sender to pass the commit phase successfully and still have two different values which it may reveal successfully in the reveal phase.

Various implementations of commitment schemes are known, each has its advantages in terms of security (i.e., binding for the receiver and secrecy for the receiver), the assumed power of the two parties etc.

Two-round commitment schemes with perfect secrecy can be constructed from any collection of claw-free permutations; see [15]. It is shown in [1] how to commit to bits with statistical security, based on the intractability of certain number-theoretic problems. Dåmgard, Pedersen and Pfitzmann [7] give a protocol for efficiently committing to and revealing strings of bits with statistical security, relying only on the existence of collision-intractable hash functions. This scheme is quite practical and we adopt it for the verifiers in our protocol. For the prover, we use a commitment scheme whose binding is information theoretic and security is computational. Such schemes can be constructed from any one-way function, see [25]. For simplicity, we simply speak of committing to and revealing bits when referring to the protocols of [7] for the verifier and [25] for the prover.

The schemes we use have the property that the receiver chooses some random string in the beginning which is later used for the commitments. It is a property of these schemes that with high probability the random choice is "good" in the sense that a polynomial number of commitments can be done with the same random choice without compromising the security of the commitment scheme. This will be used for the resettable interactive proof in Section 8. For all protocols, we need to say that the properties of the commitment schemes hold in the concurrent setting.

**Claim 2.3** *(Robustness of bit commitment schemes to a concurrent setting):*

1. *The binding property of any bit commitment scheme holds in the concurrent setting.*

2. *If the committer commits only on strings chosen uniformly at random from $\{0,1\}^k$, then the secrecy property of the commitment scheme holds also in the concurrent setting.*

**Proof:** By definition, the binding property must be robust to asynchronous composition. Otherwise, the committer may play a mental game in which his real stand-alone commitment is part of an asynchronous game which he simulates, and then defeat the binding property in the normal stand-alone world.

As for the secrecy, a similar argument may be more complicated, since the receiver cannot simulate the behavior of the committer. Specifically, the committer has some information that the receiver does not have: the value of the committed string, which may be used in the other commitments. However, in our case, the committer commits only on uniformly chosen random strings. Thus, if the committer follows the protocol, then the receiver is able to simulate the rest of the environment and the above argument holds for secrecy as well. □

In the protocol we present the committers always commit on uniformly chosen random strings in $\{0,1\}^k$. Thus, the commitment scheme is secure in the concurrent setting.

## 2.3 Witness Indistinguishability

Witness indistinguishable proofs were presented in [14]. The motivation was to provide a cryptographic mechanism whose notion of security is similar though weaker than zero-knowledge, it is meaningful and useful for cryptographic protocols, and the security is preserved in an asynchronous composition. A witness indistinguishable proof is a proof for a language in NP such that the prover is using some witness to convince the verifier that the input is in the language, yet, the view of the verifier in case the prover uses witness $w_1$ or witness $w_2$ is polynomial time indistinguishable. Thus, the verifier gets no knowledge on which witness was used in the proof. The formal definition follows. For further discussion and motivation the reader is referred to [14].

We say that a relation $R$ is polynomial time if there exists a machine that given $(x, w)$ works in polynomial time in $|x|$ and determines whether $(x, y) \in R$. For any NP language there exists a polynomial time relation $R_L$ such that $L$ can be described as $L = \{x : \exists y, \quad R_L(x, y)\}$.

**Definition 2.4** *A proof system $(P, V)$ is witness indistinguishable over a polynomial time relation $R$ is for any $V'$, any large enough $x$, any $w_1, w_2$ such that $(x, w_1) \in R$ and $(x, w_2) \in R$, and for any auxiliary input $y$ for $V'$, the view of $V'$ in the interaction with $P(x, w_1)$ is polynomially indistinguishable from the view of $V'$ in the interaction with $P(x, w_2)$.*

It is shown in [14] that witness indistinguishability is preserved with asynchronous composition of proofs. More precisely, if a proof is auxiliary-input witness indistinguishable and if the prover can run in polynomial time given the witness, then asynchronous (concurrent) composition of such proofs remains auxiliary-input witness indistinguishable (with efficient provers). See [14, 11] for more details. Note that any (auxiliary-input) zero-knowledge proof (even one that is not concurrent) is also (auxiliary-input) witness indistinguishable. In our proof system, we employ a constant round (auxiliary-input) zero-knowledge proof for the languages in NP in which the prover has an efficient procedure when given a witness to the NP-assertion such as in [16]. It follows that this proof is also (auxiliary-input) witness indistinguishable. Furthermore, since such witness indistinguishable proofs are also concurrent witness indistinguishable proof, we get that the sub-proof we employ is concurrent witness indistinguishable.

## 2.4 Blackbox simulation

The initial definition of zero-knowledge [20] requires that for any probabilistic polynomial time verifier $\hat{V}$, a simulator $S_{\hat{V}}$ exists that simulates $\hat{V}$'s view. Oren [27] proposes a seemingly stronger, "better behaved" notion of zero-knowledge, known as *black-box* zero-knowledge. The basic idea behind black box zero-knowledge is that instead of having a new simulator $S_{\hat{V}}$ for each possible verifier, we have a single probabilistic polynomial time simulator $S$ that interacts with each possible $\hat{V}$. Furthermore, $S$ is not allowed to examine the internals of $\hat{V}$, but must simply look at $\hat{V}$'s input/output behavior. That is, it can have conversations with $\hat{V}$ and use these conversations to generate a simulation of $\hat{V}$'s view that is computationally indistinguishable from $\hat{V}$'s view of its interaction with $P$.

At first glance, the limitations on $S$ may seem to force $S$ to be as powerful as a prover. However, $S$ has important advantages over a prover $P$, allowing it to perform simulations in probabilistic polynomial time. First, it may set $\hat{V}$'s coin tosses as it wishes, and even run $\hat{V}$ on different sets of coin tosses. More importantly, $S$ may conceptually "back up" $\hat{V}$ to an earlier point in the conversation, and then send different messages. This ability derives from $S$'s control of $\hat{V}$'s coin tosses; since $\hat{V}$ otherwise operates deterministically, $S$ can rerun it from the beginning, exploring different directions of the conversation by trying various messages.

Indeed, all known proofs of zero-knowledge construct black-box simulations. There is no way known to make use of a verifier's internal state, nor to customize simulators based on the description of $\hat{V}$ other than by using it as a black box.[1] Thus, given the current state of the art, an impossibility result for black-box zero-knowledge seems to preclude a positive result for the older definitions of zero-knowledge.

## 2.5 Concurrent zero-knowledge

Informally, we consider a proof system that has many copies running. Each of them consists of a prover and a verifier. We require that the provers do not cooperate. This can be also thought of as one prover with many sessions, but the behavior of the prover in any of the proof does not depend only on the current proof and not on the other copies of the proof. (This complicates the construction of the proof system, since there is no central control that checks whether the verifiers are trying to cheat, or that tries to coordinate the timing of various copies of the proof.) On the other hand, we allow the verifiers to coordinate their strategies and information. Following [9], we consider a setting in which a polynomial time adversary controls many verifiers simultaneously. The adversary $\mathcal{A}$ takes as input a partial conversation transcript of a prover interacting with several verifiers concurrently, where the transcript includes the local times on the prover's clock when each message was sent or received by the prover. The output of $A$ will be a tuples of the form $(V, \alpha, t)$, indicating that $P$ receives message $\alpha$ from a verifier $V$ at time $t$ on $P's$ local clock. The adversary may either output a new tuple as above, or wait for $P$ to output its next message to one of the verifiers. The time that is written by the adversary in the tuple, must be greater than all times previously used in the system (by messages sent to $P$ or by $P$). The view of the adversary on input $x$ in such an interaction (including all messages and times, and the verifiers random tapes) is denoted $(P, \mathcal{A})(x)$. The following definition formalizes the above using the black-box formulation. We denote by $(P, A)(x)$ the distribution on the view of $A$ in its interaction with $P$.

---

[1]As one slight exception, [22] proves security against space-bounded verifiers by considering the internal state of the verifiers. However, these techniques do not seem applicable to more standard classes of verifiers.

**Definition 2.5** *We say that an interactive proof (or argument) system $(P, V)$ for a language $L$ is (computational)* concurrent zero-knowledge *if there exists a probabilistic polynomial time oracle machine $S$ (the simulator) such that for any probabilistic polynomial time adversary $\mathcal{A}$, the distributions $(P, A)(x)$ and $S^{\mathcal{A}}(x)$ are computational indistinguishable over the strings that belong to the language $L$.*

In what follows, we will usually refer to the adversary $\mathcal{A}$ as the *adversarial verifier $V^*$* or just the *verifier $V^*$*.

**Simplifications:**

To simplify the analysis, we will only care about the order of the messages sent in the interaction and not about the delays between messages. To make the output of the simulator contain the delays dictated by the verifier, our simulator may write the output with the required delays on a working tape (instead of actually writing to the output tape) and after the order of messages and their delays has been determined and written (by the simulator we describe in this paper), the simulator may run a final stage in which it outputs the messages written on the working tape, in the same order and while inserting the appropriate delays between the messages.

We further simplify the analysis by assuming that the prover responds immediately and with no delay to the verifier's messages. Of course, in real life, the adversarial verifier may insert a few quick messages before the prover outputs his message. However, this has no effect on our proof. The rewinding strategy should be modified to consider only verifier messages. Two points should be noted: the simulator may be easily adapted to output the prover's message at times dictated by the verifier rather than immediately. The running time remains polynomial. The probability that the simulator is able to finish the simulation process after "solving" all protocols (see Section 5 below) does not change. Note that within one protocol this changes nothing. The order of messages remain the same and cannot be changed within one protocol.

## 2.6 The complexity parameters

In this paper, we simplify the discussion by using a single security parameter $k$. Our proof has $\omega(\log^2 k)$ rounds and the security is preserved with a polynomial (in $k$) number of concurrent proofs. It is possible to separate the number $k$ of allowed concurrent proofs from the security parameter. If we know that the number of proofs to be run concurrently is substantially smaller than the security parameter, then the number of rounds relates (poly-logarithmically) to the number of possible proofs (and not length of the input or the security parameter).

# 3 Main result

Our main result is the existence of poly-logarithmic round concurrent (and black box) zero-knowledge interactive proof for NP. This result builds on the following assumptions.

**Cryptographic Assumptions:** We assume the existence of two rounds commitment schemes with statistical security and the existence of two rounds commitment schemes with statistical binding.

Both assumptions are implied by the existence of a family of claw-free permutation pairs (see Section 2.2)

Let us state our main theorem given the above assumption.

**Theorem 3.1** *Assume the above cryptographic assumptions. Let $k$ be a complexity parameter bounding the size of the input. The verifier is polynomial time in $k$, and the concurrent proof may contain a polynomial (in $k$) number of proofs concurrently. Then for any function $t(\cdot)$ asymptotically greater than $\log^2 k$, i.e., $t(k) = \omega(\log^2 k)$, there exists a $t(k)$-round zero-knowledge interactive proof for all languages in NP which is: computational, black-box, and concurrent.*

This theorem is proven in Sections 4, 5, 6, and 7. First we present the protocol, next we present the simulator, and last, we analyze the simulator. Note that the proof transforms any round efficient zero-knowledge interactive proof in which the prover can be implemented in polynomial time when given access to a witness to the NP theorem proven, into a concurrent zero-knowledge interactive proof.

For resettable zero-knowledge, the construction takes as a starting point the round-efficient proof of Goldreich and Kahan [16] and uses its specific structure. The theorem about round-efficient resettable zero-knowledge is proven in Section 8:

**Theorem 3.2** *Assume the above cryptographic assumptions. Let $k$ be a complexity parameter bounding the size of the input. The verifier is polynomial time in $k$, and thus, may initiate at most a polynomial number (in $k$) of incarnations of the prover. Then for any function $t(\cdot)$ asymptotically greater than $\log^2 k$, i.e., $t(k) = \omega(\log^2 k)$, there exists a zero-knowledge argument for all languages in NP which is: computational, black-box, and resettable.*

## 4 The zero-knowledge proof system

Our concurrent zero-knowledge interactive proof system follows the ideas presented by Feige, Lapidot and Shamir [12]. On an input theorem $T$, the proof consists of a proof-preamble of $2m = \omega(\log^2 k)$ rounds and a proof body being a "standard" constant round auxiliary-input zero-knowledge proof for a modified NP theorem $T'$. The parameter $m$ determines the round complexity of the proof system. Our analysis shows that the proof system is concurrent zero-knowledge when $m$ is set to any function $m = \omega(\log^2 k)$, where $k$ is the security parameter. In the proof-body one may use any round-efficient auxiliary-input zero-knowledge proof for NP. The proof system has to be of negligible error, of round complexity smaller than $m$, and in which the prover can run in polynomial time given a witness to the NP theorem $T'$ that it must prove. All known constant round protocol such as [13, 2, 16] will do. All these (auxiliary-input) zero-knowledge proofs are also (auxiliary-input) witness indistinguishable, which is enough for us. Feige and Shamir showed that witness indistinguishability is preserved in this case also in the asynchronous setting [14]. Note that the round complexity of the resulting proof system is $O(m)$, and our proof holds when $m$ is set to any function satisfying $m = \omega(\log^2 k)$.

Let us concentrate now on the preamble, which is the main tool in transforming a regular zero-knowledge proof into a concurrent one. Let $T$ be the NP statement that the original prover would like to prove. We use a preamble with $2m$ rounds to start the proof. In this preamble, $P$ and $V$ each picks $m$ strings in $\{0,1\}^k$ denoted $p_1, p_2, \ldots, p_m$ and $v_1, v_2, \ldots, v_m$ respectively. The prover $P$ then proves that either $T$ is true or for some $i$, $1 \leq i \leq m$, $v_i = p_i$. We denote this modified theorem $T'$. For each $i$, $1 \leq i \leq m$, $P$ will have to determine $p_i$ before $v_i$ is revealed. Thus, this preamble will not give $P$ a meaningful advantage in proving the theorem. However, the simulator will be able to learn $v_i$, and then rewind the proof and set $p_i = v_i$. Thus, the simulator will have a witness to the modified theorem $T'$, and it may act as a real prover in the body of the proof. The full algorithm of the simulator is specified in Section 6 below.

The concurrent zero-knowledge argument for an input theorem $T$ goes as follows:

$$V \to P : \text{Commit to } v_1,\ v_2,\ \ldots, v_m$$
$$P \to V : \text{Commit to } p_1$$

$$V \to P : \text{Reveal } v_1$$
$$P \to V : \text{Commit to } p_2$$
$$\ldots$$
$$V \to P : \text{Reveal } v_i$$
$$P \to V : \text{Commit to } p_{i+1}$$
$$\ldots$$
$$V \to P : \text{Reveal } v_m$$
$$P \leftrightarrow V : \text{A zero-knowledge proof that } T \text{ is true or } \exists i \text{ s.t. } v_i = p_i.$$

In words: The verifier begins by committing to all its strings $v_1, \ldots, v_n$. After that, the prover commits to $p_i$ and then the verifier reveals $v_i$ for each $i$, $i = 1, 2, \ldots, m$. Finally, the prover gives a zero-knowledge proof that $T$ is true or there exists an $i$ s.t. $v_i = p_i$.

If the verifier fails to open one of its commitments properly, then the prover immediately aborts the proof. Ignoring the negligible chance that the commitments of the verifier turn out to fail the binding property, the strings $v_1, \ldots, v_m$ are fixed after the first round for the rest of the proof. Note that $v_i$ is revealed only after the prover $P$ commits on the value of $p_i$. Thus, if the security of the bit commitment holds, then $P$ can fix $p_i = v_i$ with a negligible probability. Furthermore, ignoring the negligible chance that the commitment of the prover is not secure, the verifier does not learn the value of any of the $p_i$'s so he can never tell whether it holds that $p_i = v_i$ for some $1 \leq i \leq m$.

Denote the probability that the prover fails to prove a true statement by *the completeness error* and the probability that the verifier accepts a false statement (when the prover uses an arbitrary strategy within its computational limits) *the soundness error*. We claim that these error probabilities are only slightly changed by the modification made to the proof.

**Claim 4.1** *If the original proof has soundness error $\varepsilon_s$ and completeness error $\varepsilon_c$ then the modified proof has completeness error at most $\varepsilon_c$, and soundness error at most $\varepsilon_s + \varepsilon$ for some negligible (in the security parameter $k$) $\varepsilon$.*

**Proof:** It is easy to see that the completeness property is not harmed by the modification. Regarding soundness, the additional advantage a prover $P^*$ may get over the original proof is the possibility to set $p_i = v_i$ for one of the rounds. We need to show that that cannot happen too often. Here, the security of the verifier's bit commitment is not enough. In order to make sure that the prover cannot cheat, we must require that the verifier's commitment is non-malleable [8]. In order to cheat, the prover does not need to know committed bit. It just needs to produce a commitment such that after the verifier opens its commitment to a certain string, the prover may open its commitment to the same string. Preventing this is exactly the issue in the non-malleability study, and one may use non-malleable commitment schemes as in [8] to make sure that the soundness property is preserved. However, it is not known how to achieve non-malleability in a constant round commitment schemes. Instead, we use the following trick to obtain non-malleability and keep the scheme efficient. The verifier commits using statistical secrecy. Thus, the committed value of the prover do not depend on the committed value (but with negligible probability). Next, the prover commits with an unconditionally binding scheme. Thus, the committed value binds the prover before it gets to see the verifier opening its commitment. Using these two schemes, the soundness holds. □

We remark that the problem is not symmetric: we do not need non-malleable commitment schemes for the prover. The reason for this asymmetry is that the prover never opens its commitments, so the verifier can only act upon the knowledge it gets from the commit stage. This information gives the verifier no advantage by the security property of the commitment scheme.

# 5 The simulator

We provide a black-box simulator to the above proof system (black-box simulation is discussed in Section 2.4). The adversarial verifier $V^*$ is given as a black box and the simulator interacts with it. This interaction will be used by the simulator to obtain a witness to the modified theorem $T'$. We assume that by the time the simulator gets to the body of the proof, it has such a witness. Thus, when simulating the main body, the simulator acts as the prover (which is an efficient algorithm given a witness to the NP theorem that has to be proven).

The simulator will succeed in "guessing" one of the $v_i$'s by rewinding steps in the preamble. (Recall that the real prover cannot rewind the verifier, and cannot get this advantage.) In particular, the simulator will rewind the verifier at several points during their interaction. If the verifier reveals $v_i$ before a rewind, and the simulator rewinds the verifier back far enough, it may change the value of $p_i$ and commit on $p_i = v_i$. Since the verifier is committed to the value $v_i$ (as of the first round of the interaction), then unless the rewind goes beyond the first round of the particular proof, the simulator need not worry that $v_i$ may change after it sets the commitment on $p_i$. Once the simulator has ensured that for some round $i$ $p_i = v_i$ in the preamble of a proof $\Pi$, we say that it has *solved* the proof $\Pi$. It can complete the rest of the simulation of $\Pi$ without further rewinding, by choosing $p_j$ arbitrarily for any $j \neq i$ and by playing the real prover in the main body of the proof $\Pi$ (recall that after solving a proof, it has a witness to the theorem $T'$ that has to be proven). We stress that the rest of the simulation requires no further rewinding. A key feature of this protocol is that any rewind in any of the $m$ rounds of the preamble suffices to solve the proof. Of course, if during a rewind the simulator is able to solve more than one proof by setting $p_j$'s of other proofs to values of $v_j$'s that were discovered during the first run of rewound interval, then the simulator does that. It always solves the maximum number of proofs it can in a rewind.

Note that rewinding one step in one proof may render irrelevant the simulation of steps in other proofs that took place in between those steps. Thus, choosing a step to rewind according to the need to solve a proof $\Pi$ is dangerous. It may lead the simulation to run an exponentially many steps as noted in [9] and proved for a set of protocols in [24]. We employ a different strategy of rewinding. We specify a fixed rewinding schedule regardless of the history of the interaction and the scheduling of the proofs so far. Running this rewinding schedule will guarantee a polynomial amount of work, so that the simulation is polynomial time. Nevertheless, whatever schedule of proofs the adversarial verifier-scheduler may use, the simulation is guaranteed to solve all proofs during their preamble with high probability.

Whenever the simulator rewinds the verifier, the second run of the rewind is the one that the simulator uses to continue the interaction. The first run is only used to get information and is then abandoned. The output is composed of the last full run, which is composed of second runs of relevant rewinds.

One of the most problematic issues in the design of the simulator is the following. During the run of the simulator, the adversarial verifier $V^*$ may choose to send inappropriate messages. For example, it may choose not to reveal a value $v_i$ that it has committed on in the first round. The run of the simulator is composed of rewinds: it executes an interaction with the verifier $V^*$, then it rewinds $V^*$ and makes a second run, in which it may set the $p_i$'s according to information on $v_i$'s obtained in the first run. When the adversarial verifier $V^*$ sends an inappropriate message for a proof $\Pi$ the simulator aborts sending messages to $V^*$ for this proof $\Pi$ (as the normal prover would have done). If that happens in the first run of a rewind it bears a bad affect: the simulator cannot solve the proof $\Pi$ after rewinding since it did not get to see the string $v_i$. However, if the verifier $V^*$ sends a bad message in the second run of a rewound interval, then the proof $\Pi$ is considered solved: the real prover aborts the interaction with the verifier in $\Pi$, and so does the simulator. This proof does not require solving since the body of the proof is not executed.

## 5.1 The adversarial scheduler uses round slots

We begin by simplifying our view of the adversarial schedule. Recall that we are running $k$ preambles, each with $2m$ rounds. Like discussed in Section 2.5, we assume, w.l.o.g., that the real prover (and so also the simulator) always answers immediately. Also, we do not care about delays imposed by the verifier. We only care about the order of the various rounds in the full interaction. Thus, we get that the adversarial verifier $V^*$ may schedule an overall number of $k \cdot m$ pairs of rounds in the preambles. When specifying the rewinding strategy, we are only interested in the schedule of the preambles. We do not care how the bodies of the proofs are scheduled and whether they are rewound. We will never need to rewind the bodies: the simulator will behave like the real prover in the bodies, however, bodies may be rewound due to requirements on preambles of other proofs that run concurrently.

Let us make a remark about the possibility that the adversarial verifier schedules messages in parallel. In the sequel, we do not explicitly consider parallel pairs of rounds. If the adversary sends more than one verifier's message to the prover in parallel, then the prover answers all of them in parallel. Thus, we get less then $k \cdot m$ pairs of actual rounds run. In the analysis we will analyze the probability that "something bad" happens within a specific proof, ignoring the rest of the proofs that run with it. Thus, it will not matter if this proof is run in parallel to other proofs. Note that rounds of the same proof cannot run in parallel, since the order within a proof is guaranteed to be preserved in the concurrent setting. Parallel repetitions will reduce the number of pairs of rounds and that may only make the simulation more efficient. We will not explicitly discuss parallel repetitions in the sequel.

For simplicity, from now on we will abuse the term *round* to denote a pair of rounds. Namely, in what follows, a round consists of a message of the verifier followed by an immediate response by the prover.

To summarize, we have reduced our view of the scheduled proofs to the adversarial verifier $V^*$ scheduling $km$ preamble rounds, with the only constraint that within a proof the order of rounds is preserved. We think of this schedule as assigning rounds of the various (preambles of) proofs to $km$ "slots" of rounds. We consider the $km$ slots by their order in time, and specify the rewinding strategy with respect to these slots, regardless of how the adversary assigns actual proof rounds to these slots. For example, we may let the simulator rewind the verifier to the first slot after running the second slot. More generally, After reading the verifier's message in any of the round-slots, the simulator may rewind the simulation (and the verifier) to any previous round-slot of the simulation. We will specify rewinding in the following manner. A rewind $(i \leftarrow j)$, for $1 \leq i < j \leq km$, means that after reading the verifier's message of round slot $j$, the simulator rewinds the verifier back to just before the prover message in round slot $i$. When running the rewound interval the second time, the simulator may change its message in round slot $i$ as well as any other message it made in the round slots between $i$ and $j - 1$.

## 5.2 Specification of the rewind timing

We use recursion to specify the rewinding timing. At the top level of the recursion, the simulator is running all the round slots $1..mk$. The simulator rewinds the first half of the round slots and then the second half of these round slots (regardless of which rounds of which proofs appear in the round-slots). It then "feeds" each of these $mk/2$ round slots to the recursion. Namely, at the second level of the recursion, each of the halves is split into halves and each quarter is rewound. In case the number of round slots is odd, we let the first half contain $\lceil mk/2 \rceil$ round slots and the second half contain $\lfloor mk/2 \rfloor$ round slots. Finally, at the bottom of the recursion there is an interval containing one round slot. There is no need to rewind one round slot (yet an interval of two round slots is rewound).

Let us explain this rewinding schedule with some examples. Suppose the number of round slots, $mk$, is 4. Then the round slots are run by the simulator in the following order: 1, 2, 1, 2, 3, 4, 3,

4. Using the rewinding syntax with the above sequence, we may write: $(1 \leftarrow 2), (3 \leftarrow 4)$. When $mk = 8$, the round slots have the following order: 1, 2, 1, 2, 3, 4, 3, 4, 1, 2, 1, 2, 3, 4, 3, 4, 5, 6, 5, 6, 7, 8, 7, 8, 5, 6, 5, 6, 7, 8, 7, 8. Using the rewinding syntax with the above sequence, we may write: $(1 \leftarrow 2), (3 \leftarrow 4), (1 \leftarrow 4), (1 \leftarrow 2), (3 \leftarrow 4), (5 \leftarrow 6), (7 \leftarrow 8), (5 \leftarrow 8), (5 \leftarrow 6), (7 \leftarrow 8)$.

## 6 Analysis of the simulator with respect to a static schedule

To simplify the presentation of the analysis, we start in this section by showing that the simulator works well for a static schedule. In a static schedule, the adversarial verifier $V^*$ chooses the (worst possible) schedule for the simulator, but this schedule is fixed and does not change during the simulation. In Section 7 below, we extend the argument to the case that the schedule is dynamic and may change as a function of the adversary's random coins and the history of the simulation so far.

We first note that the overall number of rounds run in the rewinding recursion, as specified, is at most $(mk)^2$ and thus, the simulator runs in polynomial time. Also, the simulator plays almost exactly the role of the prover in the second run of all rewinds (which also include the output of the simulation). There are two differences. One is that some of the committed values $p_i$ satisfy $p_i = v_i$. However, if $\hat{V}$'s behavior is changed because of this fact in a way that is noticeable in polynomial time, then the commitment scheme of the prover are not secure. The other change in the protocol is that the simulator uses different witnesses than the prover normally uses in the body of the protocol. But that difference is polynomially indistinguishable since the proof body is witness indistinguishable (also in the concurrent setting). So $\hat{V}$'s behavior and view in the simulation is polynomial time indistinguishable from its behavior and view when interacting with the real prover (both with respect to the content of the messages and to their schedule). To summarize, the output of the simulation is indistinguishable from the interactive proof assuming that the simulator solves all copies of the proof.

Our goal is to show that with overwhelming probability the simulator will manage to obtain a witness for $T'$ during the simulation of the preamble. We start with some properties of the rewinding schedule. We denote the intervals that are rewound *rewind intervals*. Because of the (recursive) manner we defined the rewinding schedule, the rewind intervals are either disjoint or contained within each other. So for any two rewind intervals $(i \leftarrow j)$ and $(k \leftarrow \ell)$ if $i < \ell \leq j$, then it holds that $k$ must be greater or equal to $i$. In the above case, in which the rewind interval $(k \leftarrow \ell)$ is contained within the rewind interval $(i \leftarrow j)$ we will say that the rewind $(i \leftarrow j)$ *dominates* the rewind $(k \leftarrow \ell)$.

**Definition 6.1** *We say that a rewind $(k \leftarrow \ell)$* dominates *the rewind $(i \leftarrow j)$ if $k \leq i < j \leq \ell$.*

We call a run of the simulator against a (black-box) verifier $V^*$ *good* if the simulator solves each of the proofs during the preamble and before it gets to simulating the main body of the proof. We would like to show that the above rewinding timing lets the simulator get "good" runs with overwhelming probability, no matter what schedule is chosen for the messages in the proofs. During the simulation, we do not need to rewind bodies of proofs, though, of course rewindings of a proof body that happens while rewinding a preamble of another proof does not hurt the simulation.

A proof $\Pi$ may be solved via a rewind $(i \leftarrow j)$ if there are at least two rounds of $\Pi$ appearing within the round slots $i, i+1, \ldots, j$, and the proof $\Pi$ does not begin or end during the round slots $i+1, i+2, \ldots, j$. The reason for precluding the first round is that if we rewind past the first round, the verifier may pick a new vector $v_1, \ldots, v_m$ and make the information obtained in the first run of the rewind useless. The proof is *actually* solved in such a rewind if the verifier behaves "well" (i.e., follows the protocol) in both runs of the rewind interval. In this case, we have two consecutive rounds of the proof $\Pi$: rounds $a$ and $a+1$ ($2 \leq a \leq m-2$) of $\Pi$ within the rewind interval. Thus, in the second run of these rounds, the simulator can set $p_a$ in the preamble to the value $v_a$ obtained in the first round and solve the proof. The reason we preclude the first round of the proof $\Pi$ from the rewind interval is that if we rewind past the first round of the proof, then $V^*$ gets to run its first round again and it may choose new values for $v_1, \ldots, v_m$. In particular, $v_a$ may change, and the simulator would not know the new value of $v_a$ to set as $p_a$. The reason that we require that the preamble does not end before the rewind, i.e., that round $m$ of the proof $\Pi$ is not within the rewind interval, is that a proof must be solved before the preamble ends. Else, the main body may start, and the simulator will noticeably fail to simulate the proof body, possibly causing the verifier to stop cooperating with the rest of the simulation.

We would like to point out that a rewind may solve the proof in any level of the recursion. If there exists a rewind $(i \leftarrow j)$ that may solve the proof, and there exists a larger rewind $(k \leftarrow \ell)$ that dominates it, then the fact that we rewind $(k \leftarrow \ell)$ does not "ruin" the solution of the proof obtained in rewind $(i \leftarrow j)$. This is true since in both runs of the rounds $\ell, \ell+1\ldots, k$ in the dominating rewind interval we rewind $(i \leftarrow j)$. So even if the rewind $(i \leftarrow j)$ happens again and again because of dominating rewinds, in each of the runs it may solve the proof again.

In what follows, we will restrict our attention to the minimal rewinding intervals that may solve a proof. If a proof may be solved by a rewind $(k \leftarrow \ell)$, then sometimes it may also be solved by several rewinds that dominate $(k \leftarrow \ell)$ just because they dominate it. However, we will be interested only in the smallest rewind interval that may solve a proof. Minimality is expressed in Conditions (1) and (4) of the following definition. This minimality property will be used to get independence between the relevant intervals.

**Definition 6.2** *We say that a rewind $(k \leftarrow \ell)$ may solve a proof $\Pi$ if the following four conditions hold:*

1. *Exactly two rounds of the preamble of $\Pi$ take place during round slots $k, k+1, \ldots, \ell$,*

2. *the first round of $\Pi$ takes place at a round slot $i < k$,*

3. *the last round of $\Pi$ takes place at a round slot $j > \ell$, and*

4. *The first round of $\Pi$ appears in the first half of the rewind interval $(k \leftarrow \ell)$ and the second round of $\Pi$ appears in the second half of the rewind interval $(k \leftarrow \ell)$.*

Note that if $\Pi$ has two rounds in the same half of the rewind interval $(k \leftarrow \ell)$, then there exists a dominated interval that may solve $\Pi$. This the ratio behind Part (4) of the definition. We now show that many rewinds may solve $\Pi$.

**Lemma 6.3** *For any schedule of $k$ copies of the proof preambles (in the $mk$ round slots), if a preamble of a specific proof $\Pi$ completes in round slot $\ell$, then there are at least $\left\lceil \frac{m}{\log(mk)+1} \right\rceil - 2$ rewind intervals that complete by round $\ell$ and that may solve $\Pi$.*

**Proof:** We first show that there are at least $\left\lceil \frac{m}{\log(mk)+1} \right\rceil$ rewind intervals that satisfy Conditions (1) and (4) in Definition 6.2 above. We then note that at most two of these intervals may foil Conditions (2) or (3), thus the number of rewinds that may solve the proof $\Pi$ is at least $\left\lceil \frac{m}{\log(mk)+1} \right\rceil - 2$ as required. Clearly, any relevant interval must end by round $\ell$, since the preamble terminates at round $\ell$.

Fix a proof $\Pi$ and any schedule of the rounds for all the proofs. We denote a rewind interval *good* if it satisfies Conditions (1) and (4) in Definition 6.2 above (with respect to $\Pi$). Consider the rewinds by the height of the recursion. At the top level, i.e., recursion height $\lceil \log(mk) \rceil$, we have $mk$ round slots. In these round slots we have $m$ rounds of the proof $\Pi$. In each recursion invocation, all round slots of the current level rewind interval are split into two almost[2] equal parts and participate in two rewind intervals of a lower recursive level. This splitting goes down the recursion until we are left with one or two round slots at recursion level 1. If we consider the rounds of the specific proof $\Pi$ as scheduled in the round slots, then there are $m$ rounds scattered at the top level, which are split into two in each recursion invocation. The split of these rounds of $\Pi$ is not necessarily equal (or even close to equal), since there may be other rounds of other proofs that appear in the (equal) split of the round slots.

---
[2]If the number of round slots is odd, then the left interval has one more round slot than the right interval.

In the following, we claim that if there are $r$ rounds of $\Pi$ in a rewind interval of level $h$, then these rounds participate in at least $\left\lceil \frac{r}{h+1} \right\rceil$ good rewind intervals with respect to $\Pi$. Assigning the recursion level $h \leq \log(mk)$ of the top level, and the number $r = m$ of rounds in the preamble of $\Pi$ in the top level, we get the validity of the assertion in Lemma 6.3.

**Claim 6.4** *For any schedule of $k$ copies of the proof (in $mk$ round slots), and for any specific proof $\Pi$. Let $r$ be an integer, $2 \leq r \leq m$, and let $h$ be an integer such that $r \leq 2^h$. Suppose there are $r$ rounds of a proof $\Pi$ in a rewind interval of recursion level $h$. Then these rounds participate in at least $\left\lceil \frac{r}{h+1} \right\rceil$ good rewind intervals with respect to the proof $\Pi$.*

**Proof:** We prove the claim by an induction on $r$. Let $r = 2$. if the two rounds are split in the current recursion invocation, then the current rewind interval is good. Otherwise, the two rounds may stay together for several invocations of the recursion and then get split, thus, making a good rewind interval at some lower level. Finally, they may stay together until the bottom level, which makes the bottom level a good rewind interval with respect to the proof $\Pi$. Thus, these 2 rounds participate in at least 1 good rewind interval, as required.

Now, suppose that the claim is correct for all $2 \leq r' < r$ and let us prove that it holds for $r$ rounds. Consider the partitioning of the $r$ rounds of the current rewind interval into two rewind intervals when invoking the next recursion. (Recall that each rewind interval is split into two rewind intervals.) Denote by $r_1$ the number of rounds that go into the first rewind interval, and by $r_2$ the number of rounds that are assigned into the second rewind interval. We know that $r_1 + r_2 = r$ and assume w.l.o.g. that $r_1 \leq r_2$. We split the analysis into 3 possible cases.

**Case 1:** $r_1 \geq 2$. In this case, we may use the induction hypothesis. The recursion level of the two rewind intervals that contain the $r_1$ and $r_2$ rounds is $h - 1$. By the induction hypothesis, the number of good rewind intervals is at least:

$$\left\lceil \frac{r_1}{h} \right\rceil + \left\lceil \frac{r_2}{h} \right\rceil \geq \left\lceil \frac{r_1 + r_2}{h} \right\rceil \geq \left\lceil \frac{r}{h+1} \right\rceil$$

and we are done with Case 1.

**Case 2:** $r_1 = 1$. In this case, we know that $r_2 = r - 1 \geq 2$ (since $r \geq 3$), thus, we may use the induction hypothesis for the second rewind interval. Nothing is guaranteed for the first rewind interval to which only one round was assigned. By the induction hypothesis, we get that the number of good rewind intervals is at least:

$$\left\lceil \frac{r_2}{h} \right\rceil = \left\lceil \frac{r-1}{h} \right\rceil \geq \left\lceil \frac{r}{h+1} \right\rceil$$

and we are done with Case 2.

**Case 3:** $r_1 = 0$. In this case, we cannot use the induction hypothesis, since $r_2 = r$. Thus, we check what may happen to these $r$ rounds as we go down the recursion. These rounds may stay together in a single rewind interval only at recursion levels greater than $\lceil \log(r) \rceil$, since there are at most $2^{h'}$ round slots at a rewind interval of recursion level $h'$. So there exists a level $2 < h' < h$ at which the rounds $r$ are split into $r_1 \geq 1$ rounds and $r_2 \geq 1$ rounds for the rewind intervals of level $h' - 1$. By the same argument as in Cases 1 and 2, we get that the number of good intervals that these $r$ rounds participate in is at least:

$$\left\lceil \frac{r}{h'+1} \right\rceil \geq \left\lceil \frac{r}{h+1} \right\rceil$$

and we are done with Case 3 and with the proof of Claim 6.4. □

As mentioned above, this also concludes the proof of Lemma 6.3 since for any proof $\Pi$ there are $m$ rounds at recursion level $\lceil \log(mk) \rceil$, and since only two of them may contain the first or last round of the preamble. □

## 6.1 Why the rewinding works

We would like to claim now that the simulator is able to solve each proof during its preamble and before it is required to simulate the main body of the proof with high probability. By Lemma 6.3, for each of the $k$ proofs, there are at least $\left\lceil \frac{m}{\log(mk)} \right\rceil - 2$ rewind intervals that may solve it. Of course, it is enough that for each proof there is one rewind that actually solves it during the preamble (rather than may solve it). If we have one such rewind for each proof, the simulator can properly simulate each proof and all of them together no matter what the schedule is.

However, it is not always the case that a proof is solved in a rewind that may potentially solve it. The reason is that the adversarial verifier $V^*$ may sometimes not open the commitment of a round of a proof $\Pi$. If the verifier $V^*$ does not open the commitment, then the real prover aborts the proof $\Pi$. In a rewind interval that may solve $\Pi$ there are exactly two rounds of $\Pi$ (which are not the first or last round). Denote the number of these rounds in the proof $\Pi$ by $a$ and $a+1$. The proof is solved in this rewind unless the following event happens:

1. the verifier *does not* reveal the committed value $v_a$ in the first run, but

2. the verifier *does* reveal the committed value $v_a$ in the second run.

All three other alternatives (i.e., the verifier reveals the committed values in both runs, or does not reveal the committed value in both runs, or reveals the committed value only in the first run) allow the simulator solve the proof $\Pi$ in this rewind. If the verifier reveals the committed value in the first run, then the proof is solved, since the simulator may set the value of its string $p_a$ to $v_a$ that it has learned. If the verifier does not reveal the committed value both in the first and second run, then the proof $\Pi$ is also solved, since the prover does not answer any of the following rounds of the proof $\Pi$, and the simulator may easily "simulate" that.

We stress that the following naive solution would not work here: output an aborted proof if either in the first or in the second run $V^*$ does not reveal the committed value. This solution is not good, since it increases the probability of aborting $\Pi$ above the probability of aborting $\Pi$ in the real proof. Thus, the simulation may become polynomially distinguishable from the original proof.

Let us compute the probability that a rewind that may solve the protocol fails to solve it. When we solve a proof, the second run is different from the first run. In particular, the value of some $p_i$ equals the value of some $v_i$ and the verifier may note that an interval is run the second time by noting that some other proof $\Pi'$ has been solved in this rewind interval. However, the prover is using a commitment scheme to secretly commit on the strings $p_i$'s in all the proofs. Using the secrecy of the commitment scheme, the verifier cannot tell that it has been rewound, so it cannot make an effort to abort the first run and behave well on the second run. Therefore, the probability that the verifier aborts in the first run is similar to the probability that it aborts in the second run of the rewind interval. These two probabilities are equal up to an (additive) negligible fraction (representing the probability that the commitment scheme fails). Whatever the probability $p$ that $V^*$ chooses not to reveal the committed value is, the probability that it does not reveal in the first run of a rewind, yet it does reveal in the second run, is $p(1 - p + \varepsilon) \leq 1/4 + \varepsilon$ for some negligible fraction $\varepsilon$. In the sequel, we assume that any rewind that may solve the proof indeed solves it with probability at least $2/3$.

We go on and compute the probability that the simulation succeeds, i.e., that each proof is solved before its preamble terminates. Note that a preamble of a proof $\Pi$ may terminate several times, since $\Pi$ may be completely (or partially) rewound several times and in particular, its last round of the preamble may be run several times. At the worst case, the preamble of each of the $k$ proofs terminates a number of times that equals the overall number of times that a rewind interval is executed. This number is at most $2^{\lceil \log(km) \rceil} \leq 2km$, i.e., a polynomial in $k$. We will show that the simulator fails to solve any particular proof with a negligible probability. Thus, it fails to solve any of (the polynomial number of) the proofs with negligible probability as well.

For any proof $\Pi$, if the preamble of $\Pi$ is completed, then the number of rounds that may solve $\Pi$ is at least $a \stackrel{\text{def}}{=} \left\lceil \frac{m}{\log(mk)} \right\rceil - 2$. Since we set $m = \omega(\log^2 k)$ and since a realistic value of $m$ satisfies $m < k$, then this number is

$$a \;\geq\; \frac{\omega(\log^2 k)}{\log(k) \;+\; \log(m)} \;=\; \omega(\log k).$$

We note now that all rewind intervals that may solve the proofs are disjoint in time. This follows since overlapping intervals must contain one another by the definition of the rewinding intervals. On the other hand, by the minimality of the intervals (Requirements (1) and (4) of Definition 6.2) a rewind that may solve the proof $\Pi$ does not contain another rewind that may solve the proof $\Pi$. Thus all these $a$ rewind intervals are disjoint and the probability that the proof is actually solved in any of them is independent and at least $1/3$. So for any occurrence of a proof $\Pi$, the probability that the simulator fails to solve it is at most $(1/3)^a$, which is a negligible fraction (in $k$). By the summation bound, the probability that the simulator fails in any of the (polynomially many) occurrences of proofs is also negligible.

## 7 Extending the analysis for the dynamic schedule

We now move to the more difficult, yet realistic case, in which the verifier does not fix the schedule of the messages in the $mk$ round slots in advance, but may determine which message to schedule in the next round slot depending on the history so far and its random coins. Looking back at the analysis of the previous section, the problem now is that the rewind intervals in which a proof may be solved constitute a random variable. Each time a new rewind interval is started, there is a probability that the interval will include two rounds of the proof (which are not the first or last round). This probability depends on the random tape of the adversarial verifier, the history so far, and the behavior of the prover (or the simulator) during the rewind interval. It is possible that in the first run of the rewind interval $V^*$ will choose to include two rounds of the proof but in the second round it will choose not to. The security of the prover's bit commitment gives us, again, a guarantee that the first run and the second run of the rewind have similar behavior.

As before, we ask ourselves what is the probability that a preamble of a proof $\Pi$ ends without the proof being solved by the simulator. At each point of the simulation one or more rewinds may start. The simulator solves the proof $\Pi$ during a rewind interval $\rho$ if the first run of $\rho$ includes exactly two rounds of the proof $\Pi$ that are not the first or the last rounds, and the verifier reveals its committed value properly. Let us present the explicit definitions.

**Definition 7.1** *(Dynamic analogue of Definition 6.2:) we say that a run of a rewind $\rho$ (either first or second run) is* interesting *with respect to a proof $\Pi$ if it includes exactly two rounds of the preamble of $\Pi$ that are neither the first nor the last round of the preamble, and each being in a different half of the rewind interval.*

**Definition 7.2** *We say that a run of a rewind $\rho$ (either first or second run) is* good *with respect to a proof $\Pi$ if it is interesting and the verifier properly reveals its commitments in the second of these two rounds. If a run is not good with respect to $\Pi$, we call it* bad *with respect to $\Pi$.*

If the first run is good with respect to a proof $\Pi$, then the proof $\Pi$ is solved (no matter what the second run is). For each run of a rewind $\rho$, depending on the history so far, there is a probability $p_\rho$, determined by the adversarial verifier, that a run of this rewind is be good with respect to $\Pi$. By the security of the prover's commitment scheme, the probability that the first run is good is equal up to an (additive) negligible fraction to the probability that the second run is good. Lemma 7.3 is similar to Lemma 6.3.

**Lemma 7.3** *In any schedule of $k$ copies of the proof (in $mk$ round slots), if a preamble of a specific proof $\Pi$ completes in round $\ell$, then there are at least $\left\lceil \frac{m}{\log(mk)+1} \right\rceil - 2$ rewind intervals that completed before round $\ell$ with a second good run with respect to $\Pi$.*

**Proof:** In the same way as in the proof of Lemma 6.3 such number of intervals must exist in any transcript that has a completed preamble in it. It remains to recall that the simulation always continues with the second run of all previous rewinds. Thus, at any point in the simulation time, all rewind intervals that have been finished have been fixed by the second run of the rewind. This means that a transcript that has a completed preamble in it must contains at least $\left\lceil \frac{m}{\log(mk)+1} \right\rceil - 2$ such rewind intervals and they all contain the second runs of all rewinds. $\square$

By Lemma 7.3, before a preamble may complete, the history must contain at least $a \stackrel{\text{def}}{=} \left\lceil \frac{m}{\log(mk)+1} \right\rceil - 2$ good second runs. However, for the proof to be completed unsolved, all the first runs of all previously completed rewinds must be bad with respect to $\Pi$. We will show that this happens with negligible probability.

**Lemma 7.4** *The probability that there exists a preamble of a proof $\Pi$ that ends well during the simulation but is not solved is negligible.*

**Proof:** We show that for any specific copy $\Pi$ of the proof whose preamble has completed, the probability that the preamble ends well, yet $\Pi$ remains unsolved is negligible. Since there is a polynomial number of proofs and each of the preambles may end a polynomial number of times, then we get that the probability that a preamble of any of the proofs remains unsolved when it ends is negligible.

Consider the run of the simulator. At each point of the simulation, one or more rewind intervals may start. At each of these points there is some probability $p$ that the run of one of the rewinds interval will be good with respect to $\Pi$. As discussed before, if the commitment scheme that the prover uses is secure, then the probability that the first run is good is equal to the probability that the second round is good up to an additive negligible fraction. We would like to compute the probability that the preamble of the proof instance $\Pi$ ends well without being solved. By Lemma 7.3, for any possible schedule of the proof instance $\Pi$, it must include at least $a$ intervals that were good in the second run with respect to $\Pi$. By our definition of a good interval, these intervals are non-overlapping (Recall the minimality condition of Definition 7.1).

We may think of the adversarial verifier as running the following stochastic experiment, which we denote *the sequential experiment*. It runs serially through tests (which are the rewinds). For the $i$th test, based on the history so far and its random tape, the adversary chooses a probability $p_i$. (This is the probability that the first run of the interval ends well. The probability that the second run ends well is at most $p_i + \varepsilon_i$ for some negligible fraction $\varepsilon_i$.) Then, with probability $(1 - p_i)(p_i + \varepsilon_i)$ the adversary wins the test, for some negligible fraction $\varepsilon_i$. (The first run is bad and the second is good.) With probability $p_i$ it looses the whole experiment (the first run is good and the simulator has solved the proof $\Pi$). In this case we say that the adversary dies. Finally, with probability $(1-p_i)(1-p_i-\varepsilon_i)$ nothing happens, i.e., the adversary neither wins nor dies (both runs are not interesting). The goal of the adversary is to win at least $a$ tests in the experiment without dying. The probability that the adversary succeeds in the sequential experiment is an upper bound on the probability that the preamble of a proof $\Pi$ ends without being solved. The reason is that when a preamble completes without being solved, all first runs must be bad and at least $a$ runs must be good. The number of tests run during the sequential experiment is $b$. In our case $b \leq 2mk$.

We now analyze the sequential experiment with parameters $a$ and $b$.

**Claim 7.5** *Let $b$ and $a$ be two positive integers such that $a < b$ and $b$ is bounded by a polynomial (in $k$). Then the probability that the adversary wins the sequential experiment with parameters $a$ and $b$ is at most $(2/3)^a$.*

**Proof:** In the sequential experiment, the adversary chooses a probability $p_i$ in each round $1 \leq i \leq b$. In each of the tests, with probability $p_i$ the adversary fails the whole experiment. With probability $(1 - p_i)(p_i + \varepsilon_i)$ it wins the $i$th test, where $\varepsilon_i$ is a negligible fraction (in $k$). With probability $(1 - p_i)(1 - p_i - \varepsilon_i)$ nothing happens and we move to the next test.

We will show that for any $\ell \geq 0$, the probability that the adversary goes from winning $\ell$ tests to winning $\ell + 1$ tests without getting killed in between, is at most $2/3$, regardless of the choice of the probabilities $p_i$'s. From that we get that the probability that the adversary wins $a$ tests without getting killed is at most $(2/3)^a$.

Suppose the adversary has won $\ell$ tests without getting killed and it is now trying to win one more. The adversary chooses probabilities $p_i$'s and runs the tests. In each test it either dies, or it wins, or nothing happens. Let $\beta$ be the number of rounds remaining before the $b$ tests of the experiment end. The probability that the adversary wins one test before it dies and before the game ends is:

$$\mu_1 \stackrel{\text{def}}{=} \sum_{t=1}^{\beta} \left( \prod_{j=1}^{t-1} (1 - p_j)(1 - p_j - \varepsilon_j) \right) \cdot (1 - p_t)(p_t + \varepsilon_t) \qquad (1)$$

To show that this probability is less than $2/3$ no matter what the choice of the $p_j$'s is, we compute the probability of a disjoint event. The event that the adversary dies before it wins the $\ell + 1$ test. (Note that there is a third disjoint event in which the adversary does not die and does not win during the remaining $\beta$ tests.) The probability of the adversary dying before winning is:

$$\mu_2 \stackrel{\text{def}}{=} \sum_{t=1}^{\beta} \left( \prod_{j=1}^{t-1} (1 - p_j)(1 - p_j - \varepsilon_j) \right) \cdot p_t \qquad (2)$$

Comparing $\mu_1$ and $\mu_2$, we see that for each term in the summation, all the factors are the same but the last. Since the $\varepsilon_i$'s are negligible (in $k$) and $\beta$ is bounded by a polynomial (in $k$), then we get that

$$\mu_1 - \mu_2 \leq \varepsilon \qquad (3)$$

for some negligible fraction $\varepsilon$. Since $\mu_1$ and $\mu_2$ represent the probabilities of disjoint events, then we also get

$$\mu_1 + \mu_2 \leq 1. \qquad (4)$$

Combining Equations 3 and 4 we get

$$\mu_1 \leq \frac{1}{2} + \frac{\varepsilon}{2} < \frac{2}{3}$$

and we are done with the proof of Claim 7.5.  □

To summarize, the probability that the preamble of any proof instance $\Pi$ ends well without being solved, is at most $\left(\frac{2}{3}\right)^a$. Recall that $a = \left\lceil \frac{m}{\log(mk)+1} \right\rceil - 2 = \omega(\log k)$ (and $b \leq (2mk)^2$), so we get that the above is a negligible fraction in $k$. Since we have at most $mk$ instances of any of the $k$ proofs, the probability that the preamble of any of these proofs ends well without being solved by our simulator is also negligible and we are done with the proof of Lemma 7.4. □

Using Lemma 7.4, we get that the simulator fails with negligible probability. Also, as in the static case, when the simulator succeeds, it outputs an interaction that is polynomially indistinguishable from the real interaction between the adversarial verifier and the real prover.

# 8 Resettable zero-knowledge

In this section, we show how to modify our interactive proof system to make it resettable zero-knowledge. We start with the definitions and proceed with the construction.

## 8.1 Definitions

We provide the definitions of resettable zero-knowledge. For more detailed discussion and motivation the reader is referred to [3].

**Definition 8.1** *An interactive proof system* $(P, V)$ *for a language $L$ is called* resettable zero-knowledge *if for any probabilistic polynomial-time adversary $V^*$ there exists a probabilistic polynomial time simulator $M^*$ so that the following two distribution ensembles are computational indistinguishable: let $t$ be a polynomial $t = poly(n)$, let each distribution be indexed by a sequence of common inputs $x_1, \ldots, x_t \in L \cap \{0,1\}^n$ and a corresponding sequence of prover's auxiliary-inputs $y_1, \ldots, y_t$.*

Distribution 1 *is defined by the following random process which depends on $P$ and $V^*$:*

1. *Randomly select and fix $t$ random tapes $\emptyset_1, \ldots, \emptyset_t$, for $P$, resulting in deterministic strategies $P^{(i,j)} = P_{x_i, y_i, \emptyset_i}$ defined by $P_{x_i, y_i, \emptyset_i}(\alpha)$ being the output of $P$ on input $x_i$ auxiliary $y_i$, random tape $\emptyset_i$ and a history so far $\alpha$, for all $i, j \in \{1, \ldots, t\}$. Each $P^{(i,j)}$ is called an incarnation of $P$.*

2. *Machine $V^*$ is allowed to run polynomially many sessions with the $P^{(i,j)}$'s. $V^*$ is allowed to send arbitrary messages to each of the of the $P^{(i,j)}$ and obtain the responses of $P^{(i,j)}$ to such messages.*

Distribution 2: *the output of $M^*(x_1, \ldots, x_t)$.*

We say that an interactive proof is black-box resettable zero-knowledge if there is a single universal simulator that can simulate an adversarial verifier $V^*$ by using $V^*$ as a black box.

As will be shown in Section 8.2, the interactive proof that we present in this paper can be easily modified into an interactive proof that has a specific structure denoted an *admissible* proof system in [3]. Canetti et. al. provide an easy way to prove that an admissible interactive proof is resettable zero-knowledge. Intuitively, an admissible proof is a proof in which the verifier restricts itself in the beginning of the protocol to its answers in the rest of the protocol. In our modified proof (described in Section 8.2 below), the verifier makes a commitment on all its future messages. In the rest of the proof system it sends its messages by revealing the committed values from the beginning of the proof. The prover aborts in case the verifier fails to reveal one of its strings appropriately.

The formal definition of admissible proofs partitions the consideration of the verifier commitments and the rest of the proof, by considering two independent prover modules. One of them deals with the commitments and verification of string revelation, and the other with the rest of the proof. Each revelation of a string contains the string itself (called the main part of the message) and a part witnessing that the revelation is correct (called the authenticator part of the message).

Let us set some conventions before continuing with the formal definition. We assume that the proof always starts with a verifier message specifying an incarnation of a prover $P^{(i,j)}$. The second message is then sent by the prover and is called the *initialization message*. (In the special interesting case the proof starts with a commitment, this message contains the random string of the prover required to make the verifier commitment). The third message is a verifier message denoted the *determining message*. We require that this message includes as a prefix the first two messages of the proof.

**Definition 8.2** *A proof system $(P, V)$ is called* admissible *if the following requirements hold:*

1. *The prover $P$ consists of two modules $P_1, P_2$. Similarly, the random input ø is partitioned into two disjoint parts $ø^{(1)}, ø^{(2)}$, where $ø^{(i)}$ is given to $P_i$. The initialization message is sent by $P_1$.*

2. *Each verifier message other than the first is first received by $P_1$ and is interpreted as consisting of two parts, called* main *and* authenticator. *$P_1$ decides whether to accept the message or to abort. If $P_1$ accepts, it forwards the main part of the message to $P_2$ who generates the next prover message.*

3. *Let $V^*$ be an arbitrary (deterministic) polynomial-size circuit representing a possible strategy for the verifier in the interactive proof $(P, V)$. Then, except with negligible probability, $V^*$ is unable to generate two different messages for some round $\ell$ that are accepted by $P_1$ with respect to the same determining message.*

We think of an incarnation of a prover $P$ in an admissible proof as having three indices $P^{(i,j,k)}$. The first index stands for the input and matching auxiliary input $x_i$ and $y_i$. The second index stands for the random tape of the first prover $ø_j^{(1)}$ and the third index stands for the random tape of the second prover $ø_k^{(2)}$.

In Section 8.2 below, we will present a modification of our protocol that is an admissible proof. It is shown in [3] that for admissible proofs it is simpler to show that the proof is resettable zero-knowledge. In particular, any admissible proof that is *hybrid* zero-knowledge is also resettable zero-knowledge. In the hybrid model of zero-knowledge the verifier is allowed to use many incarnations of provers $P^{(i,j,k)}$ but with a restriction: No two different incarnations may have the same $k$. Namely, the verifier may run the prover many times, but the random coins of the second prover must be random and independent in each run.

Let us define hybrid zero-knowledge. In Section 8.2 below we will show that our modified proof is both admissible and hybrid zero-knowledge and therefore it is also resettable zero-knowledge. Let us define hybrid zero-knowledge and quote the related assertion from [3].

**Definition 8.3** *Let $(P, V)$ be an admissible interactive proof with $P_1, P_2$ being the two modules of the prover as in Definition 8.2. We say that $(P, V)$ is hybrid zero-knowledge if Definition 8.1 holds when $V^*$ in Distribution 1 of Definition 8.1 is restricted to interacting with incarnations of provers $P^{(i,j,k)}$ such that no two incarnations $P^{(i,j,k)}$ and $P^{(i',j',k')}$ satisfy $k = k'$.*

In Section 8.2 We will use the following corollary of [3]:

**Corollary 8.4** *[3], Corollary 9: If an admissible proof $(P, V)$ is hybrid zero-knowledge then $(P, V)$ is resettable zero-knowledge.*

## 8.2 Constructing resettable zero-knowledge

Let us now show how to modify our interactive proof system to make it resettable zero-knowledge. Our proof is an easy extension of the proof in [3]. The details are provided for self containment. We make the following two modifications. First, we fix the body of the proof to use the zero-knowledge proof for 3-Colorability by Goldreich and Kahan [16]. Recall that in the body of the proof the prover provides a zero-knowledge proof that the input theorem $T$ is true or $\exists i$ s.t. $v_i = p_i$. (See Section 4 above.) Since the original theorem $T$ is an NP statement then so is the assertion "$T$ is true or $\exists i$ s.t. $v_i = p_i$", and thus, given the preamble and the theorem $T$, it can be reduced to Graph 3-Colorability, and proven in zero-knowledge via the Goldreich-Kahan proof system. A property of this proof, is that the verifier commits on its queries, which are edges in the graph. (For the honest verifier these are edges chosen uniformly at random.)

The second modification we make is that we modify the verifier to commit on its queries in the Goldreich-Kahan proof system in the beginning of the protocol preamble together with the commitments on the values $v_1, \ldots, v_m$. Since the graph is not known at this time of the protocol (recall that the reduction uses the messages sent by the parties during the preamble), the verifier commits to $n^4$ random pairs of vertices in the graph. Once the graph is determined, the proof will consider only pairs of vertices that are edges in the graph. We remark that the soundness is not substantially changed since the commitments of the verifier are done with statistical secrecy, and since the probability to end up with a quadratic number of edge-challenges is overwhelming.

We would like to show that the resulting protocol is resettable zero-knowledge. But by Corollary 8.4, if an admissible proof $(P, V)$ is hybrid zero-knowledge then $(P, V)$ is resettable zero-knowledge. We will first argue that this protocol is admissible, and then that it is hybrid zero-knowledge, and we are done.

To show that the modified protocol is admissible, we have to define the two provers. We define the first prover module to handle the verifier's commitments. It supplies the random string required for the verifier commitments and then handles the later revelations of committed values (with no need for more random coins). The second prover module proceeds with the preamble only if the first prover accepts the revelations. Then, during the body of the proof, the second prover gets the revealed challenges and answers them. It can be easily verified that this satisfies the definition of admissible proofs.

To show that the protocol is hybrid zero-knowledge we must modify the simulator to handle the modifications in the protocol, and be able to deal with multiple incarnations of the prover using (perhaps) the same random tape for the first prover's module. We first note that the modifications in the protocol do not interfere with the simulation as described. The simulator acts as a receiver in the proof system and that can be done in polynomial time. "Solving a proof" carries the same meaning as before, namely, setting a value $p_i$ to equal $v_i$ for some preamble round $i$.

The more interesting change is the additional power of the verifier, who may ask that the first module of the prover be used with the same random string in several interactions. In terms of efficiency, this bears no effect on the simulator. The simulator can keep a list of the random coins used so far in the first prover module (i.e., to receive commitments from the verifier) and use these strings repeatedly upon request from the verifier. But will the output still be of (almost) the same distribution as the real interaction?

The distribution of the commitments as played by the first prover module is perfectly simulated since the simulator plays exactly the same algorithm as the prover. It remains to check that the influence of this power given to the verifier is not destructive to the behavior of the simulator with respect to the second prover module. To see that it is not, one must verify that the secrecy and binding properties of the commitment scheme still hold even when the receiver must receive various commitments with the same random string. Looking into the commitment scheme with perfect secrecy in [15] and [7] it turns out that this is a property of these commitment schemes. Namely, the receiver starts by choosing a random string (or a random succinct description of a function) and sends his choice to the verifier. If that choice is good (and that happens with overwhelming probability) then the committer can use that to make a polynomial number of commitments with secrecy and binding properties preserved. (Actually, the same goes also for the computational secrecy commitment scheme in [25], but that is not relevant to the argument.)

One additional issue, that has not been discussed, is whether the simulator may simulate several proofs that do not have the same input. To see that this bears no additional difficulty on the simulator, note that the input is only relevant for the body of the proof and there, the simulator behaves like the real prover, regardless of which specific input the proof uses.

Thus, we get that the modified protocol is resettable zero-knowledge, and we are done with the proof of Theorem 3.2.

## 9 Acknowledgment

We thank Cynthia Dwork, Uri Feige, Oded Goldreich, Yuval Rabani, and Amit Sahai for many illuminating conversations on this subject.